\documentclass[a4paper,12pt, epsfig]{article}
\usepackage{epsfig}
\usepackage{amssymb}
\usepackage{amsfonts}
\usepackage{amsmath}
\usepackage{graphicx}
\newskip\humongous \humongous=0pt plus 1000pt minus 1000pt

\newif\ifdtup

\catcode`\@=11

\@addtoreset{equation}{section}
\def\theequation{\thesection.\arabic{equation}}

\def\@normalsize{\@setsize\normalsize{15pt}\xiipt\@xiipt
\abovedisplayskip 14pt plus3pt minus3pt%
\belowdisplayskip \abovedisplayskip
\abovedisplayshortskip \z@ plus3pt%
\belowdisplayshortskip 7pt plus3.5pt minus0pt}

\def\small{\@setsize\small{13.6pt}\xipt\@xipt
\abovedisplayskip 13pt plus3pt minus3pt%
\belowdisplayskip \abovedisplayskip
\abovedisplayshortskip \z@ plus3pt%
\belowdisplayshortskip 7pt plus3.5pt minus0pt
\def\@listi{\parsep 4.5pt plus 2pt minus 1pt
     \itemsep \parsep
     \topsep 9pt plus 3pt minus 3pt}}

\relax

\catcode`@=12

\catcode`\@=11

\def\section{\@startsection{section}{1}{\z@}{3.5ex plus 1ex minus
   .2ex}{2.3ex plus .2ex}{\large\bf}}

\def\thesection{\arabic{section}}
\def\thesubsection{\arabic{section}.\arabic{subsection}}

\def\appendix{\setcounter{section}{0}
 \def\thesection{Appendix \Alph{section}}
 \def\thesubsection{\Alph{section}.\arabic{subsection}}
 \def\theequation{\Alph{section}.\arabic{equation}}}
\def\SymBoxes#1#2#3#4{\newdimen\un@t \un@t#3%
\raisebox{#1}{\rule{#2\un@t}{#4}\hskip-#2\un@t% lower horizontal
\@tempdimb\un@t \advance\@tempdimb by-#4\@tempcntb#2\relax%
\@whilenum{\@tempcntb>0}\do{%                         % #2 vertical lines
\rule{#4}{\un@t}\hskip\@tempdimb \advance\@tempcntb by\m@ne}%
\hskip-#2\un@t \rule[\un@t]{#2\un@t}{#4}%
\rule[\un@t]{#4}{#4}\hskip-#4%             % upper horizontal line
\rule{#4}{\un@t}}\hskip-#4}                % rightest vertical line
\begin{document}
%\begin{letter}{~}

%%%%%%Define some new commands and  macros
\newcommand{\beq}{\begin{equation}}
\newcommand{\eeq}{\end{equation}}
\newcommand{\bea}{\begin{eqnarray}}
\newcommand{\eea}{\end{eqnarray}}
\newcommand{\beas}{\begin{eqnarray*}}
\newcommand{\eeas}{\end{eqnarray*}}
\newcommand{\defi}{\stackrel{\rm def}{=}}
\newcommand{\non}{\nonumber}
\newcommand{\bquo}{\begin{quote}}
\newcommand{\enqu}{\end{quote}}
%%%%%%%%%%%%%%%%
\renewcommand{\(}{\begin{equation}}
\renewcommand{\)}{\end{equation}}
%%%%%%%%%%%%%%%%%%%%%%%%%%%%%%%%%% definitions
\def \eqn#1#2{\begin{equation}#2\label{#1}\end{equation}}
\def\IZ{{\mathbb Z}}
\def\IR{{\mathbb R}}
\def\IC{{\mathbb C}}
\def\IQ{{\mathbb Q}}
\def\de{\partial}
\def\Tr{ \hbox{\rm Tr}}
\def\H{ \hbox{\rm H}}
\def\HE{ \hbox{$\rm H^{even}$}}
\def\HO{ \hbox{$\rm H^{odd}$}}
\def\K{ \hbox{\rm K}}
\def\Im{ \hbox{\rm Im}}
\def\Ker{ \hbox{\rm Ker}}
\def\const{\hbox {\rm const.}}
\def\o{\over}
\def\im{\hbox{\rm Im}}
\def\re{\hbox{\rm Re}}
\def\bra{\langle}\def\ket{\rangle}
\def\Arg{\hbox {\rm Arg}}
\def\Re{\hbox {\rm Re}}
\def\Im{\hbox {\rm Im}}
\def\exo{\hbox {\rm exp}}
\def\diag{\hbox{\rm diag}}
\def\longvert{{\rule[-2mm]{0.1mm}{7mm}}\,}
\def\a{\alpha}
\def\dag{{}^{\dagger}}
\def\tq{{\widetilde q}}
\def\p{{}^{\prime}}
\def\W{W}
\def\N{{\cal N}}
\def\hsp{,\hspace{.7cm}}
\newcommand{\C}{\ensuremath{\mathbb C}}
\newcommand{\Z}{\ensuremath{\mathbb Z}}
\newcommand{\R}{\ensuremath{\mathbb R}}
\newcommand{\rp}{\ensuremath{\mathbb {RP}}}
\newcommand{\cp}{\ensuremath{\mathbb {CP}}}
\newcommand{\vac}{\ensuremath{|0\rangle}}
\newcommand{\vact}{\ensuremath{|00\rangle}                    }
\newcommand{\oc}{\ensuremath{\overline{c}}}
\begin{titlepage}
\begin{flushright}
SISSA 24/2010/EP
%ULB-TH/mm-dd\\
%hep-th/yymmnnn\\
\end{flushright}
%\bigskip
\def\thefootnote{\fnsymbol{footnote}}

\begin{center}
{\Large {\bf
Hidden Conformal Symmetries\\
%{\em or} \\
\vspace{0.35cm}
%and \\
%\vspace{0.3cm}
of Five-Dimensional Black Holes
}}
\end{center}

\bigskip
\begin{center}
{\large  Chethan KRISHNAN\footnote{\texttt{krishnan@sissa.it}}}% and {\large  John WANG$^2$\footnote{\texttt{jwang@niagara.edu}}}
\\
\end{center}

\renewcommand{\thefootnote}{\arabic{footnote}}

\begin{center}
{\em  { SISSA, Via Beirut 2-4,\\
I-34151 , Trieste, Italy\\
\vskip .4cm}}

\end{center}

\noindent
\begin{center} {\bf Abstract} \end{center}
Recently it was shown by Castro, Maloney and Strominger (CMS) that 4D Kerr black holes have a ``hidden" conformal symmetry. Using some old results of Cvetic and Larsen, I show that this result is very likely to hold also for the most general black holes in five dimensions arising from heterotic/type II string theory.  In particular, we show how the wave equation in these geometries in the ``near region" can be written in terms of $SL(2,\IR) \times SL(2,\IR)$ Casimirs. For the special case when the black hole has two spins but no $U(1)$ charges, detailed matches for entropy and  absorption cross sections between CFT and geometry are found. The black holes we consider need not be close to extremality.

\begin{center}
%{ {\footnotesize KEYWORDS}}:
\end{center}

%\begin{center}
%\vspace{1.6 cm}
\vfill

\end{titlepage}
\bigskip

\hfill{}
\bigskip

\tableofcontents

%\newpage

\setcounter{footnote}{0}
%\section{\bf Introduction}
\section{Introduction}

Castro, Maloney and Strominger (CMS) have given evidence in a recent paper \cite{CMS} that the physics of (even far from extremal) Kerr black holes might be captured by a conformal field theory. This is interesting because unlike most of the things that string theorists are usually fascinated by, Kerr black holes have the virtue of being real.

The evidence presented in \cite{CMS} for the ``hidden" conformal symmetry of Kerr came from three (possibly related) arguments. The first was the observation that the wave equation around the black hole in a specific limit (``near-region") could be obtained as the quadratic Casimir of  an $SL(2,\IR) \times SL(2,\IR)$ algebra. The second was a computation of the entropy of the black hole by an (unjustified, but impressively successful) extrapolation of the central charge computed in the (near-)extremal case. Finally, CMS observed that the absorption cross section for scalars in the near-region can be re-interpreted as a standard finite temperature CFT absorption cross section.

The CMS result, precisely because it is tantalizing, begs us to investigate its robustness. This paper grew out of an attempt to see whether the entropy computation by means of extrapolating the central charge was a reasonable thing for black holes in general. We consider black holes in five dimensions, which have more freedom and therefore more chances of going ``wrong". In fact, we consider the most general black holes in 5D that arise as compactifications of superstring theory, and discuss their hidden conformal structure. We find that indeed, the results of CMS seem quite robust in this context. We find all three pieces of evidence for these black holes as well, to varying degree of generality. The $SL(2,\IR) \times SL(2,\IR)$ Casimir structure of the near-region wave equation is universal for these black holes. The entropy computation and the absorption cross section, we undertake for a more restrictive class of cases where the $U(1)$ charges are zero. This case is still highly non-trivial. Adopting the CMS philosophy of extrapolating from the extremal case, we {\em assume} that there is a CFT with identical left and right right central charges\footnote{In the CMS context, there was previous evidence from the computation of the near-extremal case that the right-moving central charge was identical to the left-moving central charge. In the cases we consider, as far as I am aware, the near-extremal cases have not been investigated. So the assumption we are making is exactly that: an assumption.}. We take the form of the central charge to be the same as in the extremal case. We find that the results match precisely with the classical Bekenstein-Hawking result.

\section{Geometry, Wave Equation and $SL(2,\IR) \times SL(2,\IR)$ Structure}

In this section, we will start by considering the most general black holes of the low energy effective action for type II/heterotic string theory, toroidally compactified down to five dimensions \cite{Youm}. These solutions are determined by the mass, two independent angular momenta, and three $U(1)$ charges. Later, when we make explicit comparisons of entropy and scattering between the geometry and the CFT, we will restrict to the case where the angular momenta are  independent, but the $U(1)$ charges are set to zero. This is what is usually referred to as a doubly spinning Myers-Perry black hole: in this case, we will be able to take advantage of various results (like central charges in certain limits) that are already known in the literature and do some detailed matches. But to show that the near-region scalar field equation can be reproduced as an $SL(2,\IR) \times SL(2,\IR)$ Casimir, we do not need to make this assumption and that is what we do in this section.

The metric is quite complex and can be written in many forms, we will merely refer the reader to the Appendix of \cite{CL}. The remarkable fact is that despite the complexity of the metric, the wave equation in the geometry reduces to a surprisingly simple form \cite{CL}:
\bea
&~&{\partial\over\partial x}(x^2-{1\over 4}){\partial\over\partial x}\Phi_0
+{1\over 4}[x\Delta\omega^2-\Lambda+M\omega^2+
\label{eq:geneq1}
\\
&+&{1\over x-{1\over 2}}
({\omega\over\kappa_{+}}-m_R {\Omega^R\over\kappa_{+}}
-m_L {\Omega^L\over\kappa_{+}})^2
-{1\over x+{1\over 2}}({\omega\over\kappa_{-}}-m_R {\Omega^R\over\kappa_{+}}
+m_L {\Omega^L\over\kappa_{+}})^2]\Phi_0 = 0~, \nonumber
\eea
where $x$ is a radial coordinate related to a more conventional Boyer-Lindquist-like coordinate by
\bea
x \equiv {r^2 - {1\over 2}(r^2_{+}+r^2_{-})\over
(r^2_{+}-r^2_{-})}~,
\label{eq:xdef}
\eea
with $r_+$ and $r_-$  the locations of the outer and inner horizons. $\Delta$ is a constant that we will never need. The $\Phi_0$ arises from the expansion of the Klein-Gordon scalar field in the black hole geometry according to
\bea
\Phi\equiv\Phi_0(r)~\chi(\theta)~
e^{-i\omega t+im_\phi\phi+im_\psi\psi}
=\Phi_0(r)~\chi(\theta)~e^{-i\omega t+im_R(\phi+\psi)+im_L(\phi-\psi)}~.
\eea
%\section{$SL(2,\IR) \times SL(2,\IR)$ Structure of the Near-Region Wave Equation}
$\Lambda$ stands for the eigenvalue of the angular Laplacian, and $m_{\phi,\psi}\equiv m_R\pm m_L$ are the two azimuthal quantum numbers. General expressions for $\kappa_+, \kappa_-, \Omega_L, \Omega_R$ can be found in \cite{CL}. We will write down $\Omega_L, \Omega_R$ in terms of the angular velocities at the horizon with respect to the angles in the geometry ($\phi$ and $\psi$) here\footnote{This differs from the equations (25, 26) in \cite{CL} by a factor of two. We believe that (25, 26) is a typo: in particular, according to footnote 3 in \cite{CL} the angular momenta also follow the {\em same} relation. This is impossible because potentials and charges should transform oppositely. We have checked that consistency of the relations (like thermodynamics) requires our choice.}:
\bea
\Omega_R=\Omega_\phi+\Omega_\psi, \ \  \Omega_L=\Omega_\phi-\Omega_\psi.
\eea
In a subsequent section, we will write explicit expressions for them in a more useful form in certain special cases. For the moment, all that matters to us is that $\kappa_- > \kappa_+$ and that the Hawking temperature of the hole is determined as
\bea
T_H\equiv \frac{1}{\beta_H}=\frac{2\pi}{\kappa_+}.
\eea

In the near-region that we are interested in $\Lambda$ will reduce to $l(l+2)$ (corresponding to the angular part being an $S^3$). This region, which is the crucial region for demonstrating the origin of the conformal structure, is defined by
\bea
\omega M \ll 1, \ \ r \ll \frac{1}{\omega}.
\eea
The scalar wave equation then reduces to
\bea
{\partial\over\partial x}\Big(x^2-{1\over 4}\Big){\partial\over\partial x}\Phi_0
+{1\over 4}\Big[%{1\over x-{1\over 2}}
\frac{({\omega\over\kappa_{+}}-m_R {\Omega^R\over\kappa_{+}}
-m_L {\Omega^L\over\kappa_{+}})^2}{x-{1\over 2}}
-%{1\over x+{1\over 2}}
\frac{({\omega\over\kappa_{-}}-m_R {\Omega^R\over\kappa_{+}}
+m_L {\Omega^L\over\kappa_{+}})^2}{x+{1\over 2}}\Big]\Phi_0 = l(l+2)\Phi_0~, \nonumber
\label{eq:geneq2}
\eea

Following  \cite{CMS}, we now show that this equation can be reproduced by the introduction of ``conformal" coordinates. The number of possible azimuthal angular coordinates (and therefore independent angular momenta) in five dimensions is two. In \cite{pope} it was noticed that at least when the black hole was extremal, each of these circles ($\phi$ and $\psi$ in our notation) gave rise to an independent chiral CFT. Away from extremality, we expect to have both left and right CFTs for each of the circles, so that is what we will try to see here. So we will look at two kinds of waves that impinge on the black hole: one where the $m_\phi$ modes are not excited and the other where $m_\psi$ are not excited. In both these cases, as we show below, we can write the wave equation above as the Casimir of an $SL(2,\IR) \times SL(2,\IR)$ algebra. We show this for the case when the modes are $\psi$-singlets, i.e., $m_\psi \equiv 0$, the other case ($m_\phi=0$) is trivially related.

Before we proceed, we emphasize that there are two distinct notions of left and right in the geometry. Firstly, there is the fact that $\phi+\psi$ and $\phi-\psi$ are the coordinates used by Cvetic and Larsen and in terms of which the entropy and other thermodynamic quantities have a nice left-right split \cite{CL}. What we discussed in the last paragraph is a distinct notion, arising from a left and right CFT for {\em each} of the two angles $\phi$ and $\psi$. This is inspired by the observation in \cite{pope} that in the extremal limit, there seems to be an effective(?) CFT that can capture the entropy from each of the circles.

In any event, we introduce the conformal coordinates
\bea
w^+&=&\sqrt{{r^2-r_+^2 \over r^2-r_-^2}}\ e^{2\pi T_R \phi - 2 K_R t} \\
w^-&= &\sqrt{{r^2-r_+^2 \over r^2-r_-^2}}\ e^{2\pi T_L \phi - 2 K_L t}\\
y&=&\sqrt{{r_+^2-r_+^2 \over r^2-r_-^2}}\ e^{\pi T_R \phi+\pi T_L \phi - (K_L+K_R) t}
\eea
where
\bea
T_R=\frac{1}{\pi}\frac{\kappa_+(\kappa_--\kappa_+)}{\Omega_R(\kappa_--\kappa_+)-\Omega_L(\kappa_++\kappa_-)}, \ \ T_L=\frac{1}{\pi}\frac{\kappa_+(\kappa_-+\kappa_+)}{\Omega_R(\kappa_--\kappa_+)-\Omega_L(\kappa_++\kappa_-)}, \label{TLR}\\
K_R=\frac{\Omega_L\kappa_+\kappa_-}{\Omega_R(\kappa_--\kappa_+)-\Omega_L(\kappa_++\kappa_-)}, \ \  K_L=\frac{\Omega_R\kappa_+\kappa_-}{\Omega_R(\kappa_--\kappa_+)-\Omega_L(\kappa_++\kappa_-)}. \hspace{0.15in}
\eea
Note  that these expressions for the left and right moving temperatures hold for all Cvetic-Youm black holes. In the next section, we will use a specific version of this expression to do a more specific check.
Following \cite{CMS}, now we may define left and right-moving vectors  by
\bea
H_1=i\partial_+~ ,   H_0=i(w^+\partial_++{1 \over 2}y\partial_y)~,
H_{-1}=i(w^{+2}\partial_++w^+y\partial_y-y^2\partial_- )~,  \\
 \bar H_1=i\partial_-~,  \bar H_0=i(w^-\partial_-+{1 \over 2}y\partial_y)~,
\bar H_{-1}=i(w^{-2}\partial_-+w^-y\partial_y-y^2\partial_+ )~.
\eea
which each satisfy the $SL(2,\IR)$ algebra
\bea
[H_0,H_{\pm 1}]=\mp i H_{\pm 1}~, \quad [H_{-1},H_{1}]=-2iH_0~.
\eea
The quadratic Casimir of this
\bea
{\cal H}^2&=\bar { \cal H}^2={1 \over 4}(y^2\partial_y^2-y\partial_y )+y^2\partial_+\partial_-~.
\eea
The crucial observation is that this Casimir, when written in terms of $\phi, t , r$, reduces to the wave equation with  $m_\psi=0$ on the black hole:
\bea
\Big[{\partial\over\partial x}\Big(x^2-{1\over 4}\Big){\partial\over\partial x}% \label{eq:geneq3} \\
-{(r_+^2-r_-^2)\over 4(r^2-r_+^2)}
\Big({\partial_t\over\kappa_{+}}+{\Omega^R+\Omega^L\over\kappa_{+}}\partial_\phi
\Big)^2 + \hspace{1.5in} \label{eq:geneq3}\\
\hspace{2in}+{(r_+^2-r_-^2)\over 4(r^2-r_-^2)}
\Big({\partial_t\over\kappa_{-}}+{\Omega^R-\Omega^L\over\kappa_{+}}\partial_\phi
\Big)^2\Big]\Phi_0 = l(l+2)\Phi_0~, \nonumber
\eea

An entirely analogous construction holds also if we consider the $\psi$ circle instead of the $\phi$ circle.

\noindent
{\bf Thermal Density Matrix:} Following the same Rindler-inspired arguments as in CMS \cite{CMS}, we now notice that the periodic identification of $\phi$ %(and $\psi$)
by $2\pi$ results in a thermal density matrix at temperature $(T_L, T_R)$. It should be noted however that in our case, we find such a structure for each of the azimuthal circles of the black hole (i.e, both for $\phi$ and $\psi$). There seems to be two $SL(2, \IR) \times SL(2,\IR)$ structures in the geometry.

In \cite{pope} it was found that in the extremal limit of rotating higher dimensional black holes, there is a chiral (say, left-moving) CFT corresponding to each direction of rotation. In the 4D Kerr case, it is known that moving away from extremality amounts to turning on a right moving sector on top of the left-moving sector that was visible at extremality. So it is natural to expect that in the higher dimensional case we are interested in, if we consider the non-extremal case (like above), there should be both a left and right CFT for each direction of rotation. The doubled $SL(2,\IR)\times SL(2,\IR)$ structure we find above corroborates precisely that expectation. It would be interesting to see how these two $SL(2,\IR)\times SL(2,\IR)$'s are related. The counting of independent quantum numbers in the geometry seems to suggest that they should not be independent, it is possible that they are related by some version of a ``T-duality"\footnote{I thank S. Sheikh-Jabbari for this comment.}.

The structure we found is generic, but we can do more concrete and detailed tests in more specific cases. In the Kerr case, it was found that when the black hole was near-extremal as opposed to extremal, the left and right CFTs had central charges which were identical and equal to
\bea
c_L=c_R=12 J. \ \ \ {\rm (for\ \ 4D\ \ Kerr)}
\eea
In \cite{CMS} it was found that assuming that this form for the central charge was preserved even far away from extremality and then using the Cardy formula with the left and right temperatures (the analogues of our eq. (\ref{TLR})) reproduced the entropy of the generic Kerr black hole. We will find that a similar strategy works even for the five dimensional black holes when the $U(1)$ charges are set to zero\footnote{The more general case will not be undertaken in this paper, partly because it seems that the computations of the central charges in the extremal case are not available in the literature. But it seems likely that the result should hold there as well.} and that the entropy matches in quite a non-trivial way.

\section{Entropy From the CFT}

In this section, we will work with the case of the black hole with two spins but no Abelian charges. There are two reasons for doing this. One is that the explicit forms of the inner and outer horizons are rather simple in this case, while permitting rather non-trivial checks for the entropy. Secondly, in \cite{pope}, the extremal limit of this black hole was investigated and the central charge was computed, so we can directly use their result.

Before we proceed, one word about the notation translation between \cite{CL} and \cite{pope}: the case that we consider now corresponds to setting $\delta_i=0$ in the notations of \cite{CL}. Then, the parameters of the black hole are related as (subscripts denote Cvetic-Larsen and Lu-Mei-Pope):
\bea
\mu_{(CL)}=2M_{(LMP)}, \ \ {l_1}_{(CL)}=a_{(LMP)}, \ \ {l_2}_{(CL)}=b_{(LMP)}.
\eea
Some care needs to be exercised in matching the two notations. Firstly, the parameter $M$ as defined in LMP is not the mass that shows up in the the equations of black hole thermodynamics (if one takes their $J$ as the thermodynamic $J$). Instead, it is given by ${\cal M}=\frac{3 \pi M}{4}$: this is explicitly stated in \cite{HHT} whose notations are the same as those of \cite{pope}. This results in a factor of $4/\pi$ difference between the extensive quantities (mass, angular momentum, entropy) used by CL and LMP. This seems to be due to a difference in the choice of units in Newton's constant. The thermodynamical relations hold in both cases (of course), but there is an overall scaling by a numerical factor. We will follow the notations of \cite{pope} consistently in this section.

The central charges of the five dimensional black hole was computed in \cite{pope} in the extremal limit. It was found for the 5D black hole (with the cosmological constant set to zero in \cite{pope}) that the $\phi$-direction of rotation corresponds to a chiral CFT with central charge
\bea
c_\phi=\frac{3}{2}\pi b (a+b)^2.
\eea
A bit of algebra using the expressions in \cite{pope} shows that in terms of the mass and angular momenta of the black hole ($M, J_\phi, J_\psi$), this can be written as
\bea
c_\phi= 6 J_\psi. \label{cent}
\eea
A similar result holds for the central charge along the $\psi$ direction.

Now we make use of the observation that in 4D Kerr, near-extremal black holes corresponded to a CFT with right and left central charges\footnote{Here the final equality is supposed to capture the equality as a function of $J$. The value of the central charge will be different in the extremal and non-extremal cases.} $c_L=c_R=12 J={c_L}_{{\rm Extremal}}$. Assuming that this result holds for far-extremal black holes lead to the successful matches in \cite{CMS}. We will {\em assume} that such a statement is true also in the case of the 5D black hole. That is, we will assume that away from extremality, the CFT along the $\phi$-circle will have both a left and right-moving piece, both with identical central charges given by (\ref{cent}). Remarkably, this assumption precisely reproduces the correct entropy.

Following \cite{CMS} we will use the Cardy formula to compute the entropy of the black hole:
\bea
S=\frac{\pi^2}{3}(c_L T_L + c_R T_R) = \frac{\pi^2}{3}(6 J_\psi)(T_L+T_R).
\eea
At this stage, we need the explicit forms of $\kappa_+, \kappa_-,\Omega_L, \Omega_R$ for the doubly spinning black hole. Using the results in \cite{CL, pope}, we can compute
\bea
\Omega_L=\frac{r_+(a-b)(r_+-r_-)}{(r_+^2+a^2)(r_+^2+b^2)},  \ \ \Omega_R=\frac{r_+(a+b)(r_++r_-)}{(r_+^2+a^2)(r_+^2+b^2)} \\
\kappa_+=\frac{r_+(r_+^2-r_-^2)}{(r_+^2+a^2)(r_+^2+b^2)}, \ \ \kappa_-=\frac{r_+^2(r_+^2-r_-^2)}{r_-(r_+^2+a^2)(r_+^2+b^2)}.
\eea
Here $r_+$ and $r_-$ are the outer and inner horizons. Note that since we are working with the simplified case with no $U(1)$ charges, the equation determining them reduces effectively to a standard quadratic equation \cite{pope}. The relations in \cite{CL} are written in terms of the explicit solutions of the roots, and we have found the following two relations very useful in translating to the \cite{pope} language:
\bea
M=\frac{(r_+^2+a^2)(r_+^2+b^2)}{2r_+^2}, \ \ r_+r_-=ab,
\eea
both of which are immediate consequences of the equation determining the horizons. Using these, after a bit of algebra, we find that the left and right moving temperatures found in the last section become
\bea
T_R=\frac{r_+-r_-}{2\pi b}, \ \ T_L=\frac{r_++r_-}{2\pi b}.
\eea
These, together with the fact that
\bea
J_\psi=\frac{\pi M b}{2}
\eea
leads to the result for the entropy
\bea
S=\frac{\pi^2(r_+^2+a^2)(r_+^2+b^2)}{2 r_+},
\eea
which agrees precisely with the gravity result presented in eq. (3.3) of \cite{pope} (we are working with cosmological constant zero). It is also to be noted that the right temperature vanishes, while the left temperature goes to the result obtained in \cite{pope} in the extremal limit ($r_+=r_-\equiv r_0$):
\bea
T_R=0, \ \ T_L=\frac{a}{\pi \sqrt{ab}}. \ \ \rm{(at \ extremality).}
\eea

\section{Absorption Cross Section}

Now we show that the absorption cross-section for scalars (with $m_\psi=0$) in the near-region of the black hole can be interpreted as arising from a CFT at the left and right temperatures presented before. Again, we start with the most general black hole. The general absorption cross section is easy enough to write down using the results in \cite{CL}, but we will only present the case when $m_\psi=0$:
\bea
P_{{\rm abs}}\sim \sinh \left(\frac{\beta_H}{2}(\omega-m_\phi \Omega_\phi)\right) \times\hspace{3.5in} \\ \nonumber
\hspace{0.2in}\times \left|\Gamma\Big(\xi-i\frac{\beta_L\omega-\beta_Hm_\phi(\Omega_\phi-\Omega_\psi)}{4\pi}\Big)\right|^2
\left|\Gamma\Big(\xi-i\frac{\beta_R\omega-\beta_Hm_\phi(\Omega_\phi+\Omega_\psi)}{4\pi}\Big)\right|^2 \label{hyper}
\eea
The symbols involved are defined as:
\bea
\beta_H=\frac{2\pi}{\kappa_+}, \ \ \beta_R=\frac{2\pi}{\kappa_+}+\frac{2\pi}{\kappa_-}, \  \ \beta_L=\frac{2\pi}{\kappa_+}-\frac{2\pi}{\kappa_-,}
\eea
In the near-region approximation that we are working with, $\xi$ in \cite{CL} simplifies to the form
\bea
\xi=l_0+1,  \ \ {\rm where} \ \ l_0 \equiv l/2.
\eea
Note that with this definition for $l_0$, the eigenvalue of the $SL(2,\IR)$ Casimir that we defined earlier takes the form $2l_0(2l_0+2)=4l_0(l_0+1)$. It is $(l_0, l_0)=(h_L,h_R)$ that we identify as the conformal weights of the field.

To proceed with the matching, we first note  that the first law of the black hole thermodynamics takes the form
\bea
T_H \delta S=\delta {\cal M}-\Omega_\phi \delta J_\phi -\Omega_\psi \delta J_\psi= \omega-m_\phi \Omega_\phi-m_\psi \Omega_\psi
\eea
From this point we restrict to the case where the $U(1)$ charges of the black hole are zero. Note that in this case, as pointed out before, the mass that shows up in the thermodynamical relation is ${\cal M}=\frac{3 \pi M}{4}$ in the notations  of \cite{pope}. It can be explicitly checked that the first law of thermodynamics holds by plugging in the expressions for the various quantities. It is customary to work with mass and angular momenta as the basic variables of the black hole when doing the variations, but we  find it more convenient to work with $r_+, a, b$ as the independent variables and match the right and left hand sides in terms of them.

The next step is to identify charges such that
\bea
\delta S= \frac{\delta E_L}{T_L}+\frac{\delta E_R}{T_R}
\eea
Since we are working with the case $m_\psi=0$, we are looking at constrained variations. The basic variations are $\delta {\cal M} = \omega , \delta J_\phi = m_\phi, \delta J_\psi =m_\psi$. So we set $\delta J_\psi = m_\psi$ to zero and then solve for $\delta r_+, \delta a, \delta b$ in terms of $\omega, m_\phi$. With these at hand we find that there are two canonical choices for $\delta E_L \equiv \omega_L$ and $\delta E_R \equiv \omega_R$ that bring (\ref{hyper}) to the form of a thermal CFT absorption cross section:
\bea
 \omega_L={(a + b) (a b + r_+^2) m_\phi\over
 2 b (a b - r_+^2)} + {(a^2 + r_+^2) (b^2 + r_+^2) \omega \over (2 b r_+^2)}, \\
 \omega_R= {(a - b) (a b - r_+^2) m_\phi \over
 2 b (a b + r_+^2)} + {(a^2 + r_+^2) (b^2 + r_+^2) \omega \over 2 b r_+^2},
\eea
and,
\bea
 \omega_L=-{(a - b) m_\phi \over
 2 b} + {(a^2 + r_+^2) (a b + r_+^2) (b^2 + r_+^2) \omega \over
 2 b r_+^2 ( r_+^2-a b)}, \\
 \omega_R=-{(a +b) m_\phi \over
 2 b} + {(a^2 + r_+^2) ( r_+^2-a b) (b^2 + r_+^2) \omega \over
 2 b r_+^2 ( r_+^2+a b)}.
\eea
Either one of these choices brings the absorption cross section to the standard thermal CFT form:
\bea
P_{\rm abs}\sim T_L^{2h_L-1}T_R^{2h_R-1}\sinh\left({\omega_L\over 2T_L}+{\omega_R \over 2 T_R}\right)\left|\Gamma(h_L+i{\omega_L \over 2\pi T_L})\right|^2\left|\Gamma(h_R+i{\omega_R\over 2\pi T_R})\right|^2~.
\eea

\section{Comments}

The results of this paper suggest that even away from extremality, each azimuthal circle gives rise to an effective (left-right) CFT for rotating black holes. A natural question is how seriously should one take these CFTs, and in particular how they are related to each other. It seems that there is some redundancy because both sets of CFTs are able to reproduce the entropy of the hole.

The fact that there is an $SL(2,\IR) \times SL(2,\IR)$ seems to suggest the presence of an auxiliary $AdS_3$ that captures the wave equation. The periodicity in the angular direction is interpreted as a spontaneous breaking of  the  $SL(2,\IR) \times SL(2,\IR)$. It would be interesting to see whether one can construct a useful notion for a Green function for Kerr by starting with the non-compact space and using some version of the method of images (see \cite{CKR}). It will be very interesting to see how far one can push the usual technology of AdS/CFT in this auxiliary space: the question of what takes the role of the boundary, stress tensors, holographic renormalization, etc. might be tractable.

A program that is likely to be of immense interest from an astrophysical point of view would be to use this hidden conformal structure (or near-conformal structure) to study the physics of stellar black holes, neutron stars and supermassive black holes such as the one in the Galactic center. There are plenty of open problems there, and it might not be so far out to hope that such an approach is of some practical promise. %is a promising avenue for research.

It would also be interesting to see whether any of this can be generalized to the study of the more general black objects that have been constructed recently in five and higher dimensions \cite{5d}. Previous work on the Kerr-CFT correspondence can be found in \cite{KerrCFT}.

\section{Acknowledgments}

I want to thank Daniel Arean, Bernard Kay,  Bindusar Sahoo and Shahin Sheikh-Jabbari for discussions/correspondence. Special thanks to Shahin for comments on a preliminary version of the manuscript. Lubos' blog also deserves an acknowledgment for pointing my attention to \cite{CMS}.

% ==========================================================================
%
%%%%%%%%%%%%%%%%%%%%%%%%%%%%%%%%%%%%%%%%%%%%%%%%%%%%%%%%%%%%%%%%%%%%%%%%%%%%
%                      REFERENCES                            %
%%%%%%%%%%%%%%%%%%%%%%%%%%%%%%%%%%%%%%%%%%%%%%%%%%%%%%%%%%%%%%%%%%%%%%%%%%%%
%\newpage
%\bibliography{metasusy}

\begin{thebibliography}{19}        %here 19 is the widest mark...
%-----Type it \bibitem[how it is marked]{how we call it}Authors,
%-----Citations are then made by \cite{how we call it} in text
%-----\bibitem without [how it is denoted] is numbered 1,2,3....


%\cite{Verlinde:2010hp}

%\cite{Ramgoolam:2004gw}

%\cite{Castro:2010fd}
\bibitem{CMS}
  A.~Castro, A.~Maloney and A.~Strominger,
  ``Hidden Conformal Symmetry of the Kerr Black Hole,''
  arXiv:1004.0996 [hep-th].
  %%CITATION = ARXIV:1004.0996;%%
%\cite{Cvetic:1997uw}

\bibitem{CL}
  M.~Cvetic and F.~Larsen,
  ``General rotating black holes in string theory: Greybody factors and  event
  horizons,''
  Phys.\ Rev.\  D {\bf 56}, 4994 (1997)
  [arXiv:hep-th/9705192].
  %%CITATION = PHRVA,D56,4994;%%

%\cite{Lu:2008jk}
\bibitem{pope}
  H.~Lu, J.~Mei and C.~N.~Pope,
  ``Kerr/CFT Correspondence in Diverse Dimensions,''
  JHEP {\bf 0904}, 054 (2009)
  [arXiv:0811.2225 [hep-th]].
  %%CITATION = JHEPA,0904,054;%%

%\cite{Cvetic:1996xz}
\bibitem{Youm}
  M.~Cvetic and D.~Youm,
  ``General Rotating Five Dimensional Black Holes of Toroidally Compactified
  Heterotic String,''
  Nucl.\ Phys.\  B {\bf 476}, 118 (1996)
  [arXiv:hep-th/9603100].
  %%CITATION = NUPHA,B476,118;%%

%\cite{Hawking:1998kw}
\bibitem{HHT}
  S.~W.~Hawking, C.~J.~Hunter and M.~Taylor,
  ``Rotation and the AdS/CFT correspondence,''
  Phys.\ Rev.\  D {\bf 59}, 064005 (1999)
  [arXiv:hep-th/9811056].
  %%CITATION = PHRVA,D59,064005;%%

%\cite{Krishnan:2010df}
\bibitem{CKR}
  C.~Krishnan,
  ``Black Hole Vacua and Rotation,''
  arXiv:1005.1629 [hep-th].
  %%CITATION = ARXIV:1005.1629;%%

%\cite{Becker:2010jj}
\bibitem{KerrCFT}
 M.~Guica, T.~Hartman, W.~Song and A.~Strominger,
  ``The Kerr/CFT Correspondence,''
  arXiv:0809.4266 [hep-th].
T.~Hartman, K.~Murata, T.~Nishioka and A.~Strominger,
  ``CFT Duals for Extreme Black Holes,''
  arXiv:0811.4393 [hep-th].
H.~Lu, J.~Mei, C.~N.~Pope and J.~Vazquez-Poritz,
  ``Extremal Static AdS Black Hole/CFT Correspondence in Gauged
  Supergravities,''
  arXiv:0901.1677 [hep-th].
  D.~D.~K.~Chow, M.~Cvetic, H.~Lu and C.~N.~Pope,
  ``Extremal Black Hole/CFT Correspondence in (Gauged) Supergravities,''
  arXiv:0812.2918 [hep-th].
A.~M.~Ghezelbash,
  %``Kerr-Bolt Spacetimes and Kerr/CFT Correspondence,''
  arXiv:0902.4662 [hep-th].
  %%CITATION = ARXIV:0902.4662;%%
  M.~R.~Garousi and A.~Ghodsi,
  %``The RN/CFT Correspondence,''
  arXiv:0902.4387 [hep-th].
  %%CITATION = ARXIV:0902.4387;%%
  K.~Hotta,
  %``Holographic RG flow dual to attractor flow in extremal black holes,''
  arXiv:0902.3529 [hep-th].
  %%CITATION = ARXIV:0902.3529;%%
  T.~Fukuyama,
  %``SO(2,d-1) Gauge Theory of Gravity in d Dimensional Spacetime and
  %$AdS_d/CFT_{d-1}$ Correspondence,''
  arXiv:0902.2820 [hep-th].
  S.~Nam and J.~D.~Park,
  %``Hawking radiation from covariant anomalies in 2+1 dimensional black
  %holes,''
  arXiv:0902.0982 [hep-th].
  B.~Chen and Z.~b.~Xu,
  %``Quasinormal modes of warped $AdS_3$ black holes and AdS/CFT
  %correspondence,''
  arXiv:0901.3588 [hep-th].
  A.~M.~Ghezelbash,
  %``Kerr/CFT Correspondence in Low Energy Limit of Heterotic String Theory,''
  arXiv:0901.1670 [hep-th].
  F.~Loran and H.~Soltanpanahi,
  %``5D Extremal Rotating Black Holes and CFT duals,''
  arXiv:0901.1595 [hep-th].
  C.~M.~Chen and J.~E.~Wang,
  %``Holographic Duals of Black Holes in Five-dimensional Minimal
  %Supergravity,''
  arXiv:0901.0538 [hep-th].
  J.~J.~Peng and S.~Q.~Wu,
  %``Extremal Kerr black hole/CFT correspondence in the five dimensional G\'odel
  %universe,''
  arXiv:0901.0311 [hep-th].
  T.~Azeyanagi, N.~Ogawa and S.~Terashima,
  %``The Kerr/CFT Correspondence and String Theory,''
  arXiv:0812.4883 [hep-th].
  H.~Isono, T.~S.~Tai and W.~Y.~Wen,
  %``Kerr/CFT correspondence and five-dimensional BMPV black holes,''
  arXiv:0812.4440 [hep-th].
  Y.~Nakayama,
  %``Emerging AdS from Extremally Rotating NS5-branes,''
  arXiv:0812.2234 [hep-th].
  A.~Garbarz, G.~Giribet and Y.~Vasquez,
  %``Asymptotically AdS3 Solutions to Topologically Massive Gravity at Special
  %Values of the Coupling Constants,''
  arXiv:0811.4464 [hep-th].
  T.~Azeyanagi, N.~Ogawa and S.~Terashima,
  %``Holographic Duals of Kaluza-Klein Black Holes,''
  arXiv:0811.4177 [hep-th].
  G.~W.~Gibbons, C.~A.~R.~Herdeiro, C.~M.~Warnick and M.~C.~Werner,
  %``Stationary Metrics and Optical Zermelo-Randers-Finsler Geometry,''
  arXiv:0811.2877 [gr-qc].
  K.~Hotta, Y.~Hyakutake, T.~Kubota, T.~Nishinaka and H.~Tanida,
  %``The CFT-interpolating Black Hole in Three Dimensions,''
  JHEP {\bf 0901}, 010 (2009)
  [arXiv:0811.0910 [hep-th]].
  M.~Schvellinger,
  %``Kerr-AdS black holes and non-relativistic conformal QM theories in diverse
  %dimensions,''
  JHEP {\bf 0812}, 004 (2008)
  [arXiv:0810.3011 [hep-th]].
  %%CITATION = JHEPA,0812,004;%%
  M.~Becker, S.~Cremonini and W.~Schulgin,
  ``Extremal Three-point Correlators in Kerr/CFT,''
  arXiv:1004.1174 [hep-th].
  J.~Mei,
  ``The Entropy for General Extremal Black Holes,''
  JHEP {\bf 1004}, 005 (2010)
  [arXiv:1002.1349 [hep-th]].
  C.~M.~Chen, Y.~M.~Huang and S.~J.~Zou,
  ``Holographic Duals of Near-extremal Reissner-Nordstrom Black Holes,''
  JHEP {\bf 1003}, 123 (2010)
  [arXiv:1001.2833 [hep-th]].
  J.~J.~Peng and S.~Q.~Wu,
  ``Extremal Kerr/CFT correspondence of five-dimensional rotating (charged)
  black holes with squashed horizons,''
  Nucl.\ Phys.\  B {\bf 828}, 273 (2010)
  [arXiv:0911.5070 [hep-th]].
  C.~Krishnan,
  ``Tomograms of Spinning Black Holes,''
  Phys.\ Rev.\  D {\bf 80}, 126014 (2009)
  [arXiv:0911.0597 [hep-th]].
  C.~M.~Chen, J.~R.~Sun and S.~J.~Zou,
  ``The RN/CFT Correspondence Revisited,''
  JHEP {\bf 1001}, 057 (2010)
  [arXiv:0910.2076 [hep-th]].
  J.~Rasmussen,
  ``A note on Kerr/CFT and free fields,''
  arXiv:0909.2924 [hep-th].
  D.~Grumiller and A.~M.~Piso,
  ``Exact relativistic viscous fluid solutions in near horizon extremal Kerr
  background,''
  arXiv:0909.2041 [astro-ph.SR].
  J.~Rasmussen,
  ``Isometry-preserving boundary conditions in the Kerr/CFT correspondence,''
  Int.\ J.\ Mod.\ Phys.\  A {\bf 25}, 1597 (2010)
  [arXiv:0908.0184 [hep-th]].
  Y.~Matsuo, T.~Tsukioka and C.~M.~Yoo,
  ``Yet Another Realization of Kerr/CFT Correspondence,''
  Europhys.\ Lett.\  {\bf 89}, 60001 (2010)
  [arXiv:0907.4272 [hep-th]].
  Y.~Matsuo, T.~Tsukioka and C.~M.~Yoo,
  ``Another Realization of Kerr/CFT Correspondence,''
  Nucl.\ Phys.\  B {\bf 825}, 231 (2010)
  [arXiv:0907.0303 [hep-th]].
  O.~J.~C.~Dias, H.~S.~Reall and J.~E.~Santos,
  ``Kerr-CFT and gravitational perturbations,''
  JHEP {\bf 0908}, 101 (2009)
  [arXiv:0906.2380 [hep-th]].
  L.~M.~Cao, Y.~Matsuo, T.~Tsukioka and C.~M.~Yoo,
  ``Conformal Symmetry for Rotating D-branes,''
  Phys.\ Lett.\  B {\bf 679}, 390 (2009)
  [arXiv:0906.2267 [hep-th]].
  W.~Kim and E.~J.~Son,
  ``Central Charges in 2d Reduced Cosmological Massive Gravity,''
  Phys.\ Lett.\  B {\bf 678}, 107 (2009)
  [arXiv:0904.4538 [hep-th]].
  X.~N.~Wu and Y.~Tian,
  ``Extremal Isolated Horizon/CFT Correspondence,''
  Phys.\ Rev.\  D {\bf 80}, 024014 (2009)
  [arXiv:0904.1554 [hep-th]].
  C.~Krishnan and S.~Kuperstein,
  ``A Comment on Kerr-CFT and Wald Entropy,''
  Phys.\ Lett.\  B {\bf 677}, 326 (2009)
  [arXiv:0903.2169 [hep-th]].
  T.~Azeyanagi, G.~Compere, N.~Ogawa, Y.~Tachikawa and S.~Terashima,
  ``Higher-Derivative Corrections to the Asymptotic Virasoro Symmetry of 4d
  Extremal Black Holes,''
  Prog.\ Theor.\ Phys.\  {\bf 122}, 355 (2009)
  [arXiv:0903.4176 [hep-th]].
  W.~Y.~Wen,
  ``Holographic descriptions of (near-)extremal black holes in five dimensional
  minimal supergravity,''
  arXiv:0903.4030 [hep-th].
  %%CITATION = ARXIV:0903.4030;%%
  %\cite{Chen:2010ni}
  B.~Chen, B.~Ning and Z.~b.~Xu,
  ``Real-time correlators in warped AdS/CFT correspondence,''
  JHEP {\bf 1002}, 031 (2010)
  [arXiv:0911.0167 [hep-th]].
  %%CITATION = JHEPA,1002,031;%%
  B.~Chen and C.~S.~Chu,
  ``Real-time correlators in Kerr/CFT correspondence,''
  JHEP {\bf 1005}, 004 (2010)
  [arXiv:1001.3208 [hep-th]].
  %%CITATION = JHEPA,1005,004;%%


\bibitem{5d}
  R.~Emparan and H.~S.~Reall,
  ``A rotating black ring in five dimensions,''
  Phys.\ Rev.\ Lett.\  {\bf 88}, 101101 (2002)
  [arXiv:hep-th/0110260].
  %%CITATION = PRLTA,88,101101;%
  A.~A.~Pomeransky and R.~A.~Sen'kov,
  ``Black ring with two angular momenta,''
  arXiv:hep-th/0612005.
  H.~Elvang and P.~Figueras,
  ``Black Saturn,''
  JHEP {\bf 0705}, 050 (2007)
  [arXiv:hep-th/0701035].
  J.~Evslin and C.~Krishnan,
  ``Metastable Black Saturns,''
  JHEP {\bf 0809}, 003 (2008)
  [arXiv:0804.4575 [hep-th]].
  H.~Iguchi and T.~Mishima,
  ``Black di-ring and infinite nonuniqueness,''
  Phys.\ Rev.\  D {\bf 75}, 064018 (2007)
  [Erratum-ibid.\  D {\bf 78}, 069903 (2008)]
  [arXiv:hep-th/0701043].
  J.~Evslin and C.~Krishnan,
  ``The Black Di-Ring: An Inverse Scattering Construction,''
  Class.\ Quant.\ Grav.\  {\bf 26}, 125018 (2009)
  [arXiv:0706.1231 [hep-th]].
  K.~Izumi,
  ``Orthogonal black di-ring solution,''
  Prog.\ Theor.\ Phys.\  {\bf 119}, 757 (2008)
  [arXiv:0712.0902 [hep-th]].
  %%CITATION = JHEPA,0804,045;%%
  %\cite{Elvang:2007hs}
  H.~Elvang and M.~J.~Rodriguez,
  ``Bicycling Black Rings,''
  JHEP {\bf 0804}, 045 (2008)
  [arXiv:0712.2425 [hep-th]].
  %%CITATION = JHEPA,0804,045;%%
  R.~Emparan and H.~S.~Reall,
  ``Black Holes in Higher Dimensions,''
  Living Rev.\ Rel.\  {\bf 11}, 6 (2008)
  [arXiv:0801.3471 [hep-th]].
  %%CITATION = 00222,11,6;%%

\end{thebibliography}

\end{document}